\documentstyle[epsf]{l-aa}

\catcode`@=11
\def\eqalign#1{\null\,\vcenter{\openup\jot\m@th
  \ialign{\strut\hfil$\displaystyle{##}$&$\displaystyle{{}##}$\hfil
      \crcr#1\crcr}}\,}
\def\eqalignleft#1{\null\,\vcenter{\openup\jot\m@th
  \ialign{\strut$\displaystyle{##}$\hfil&$\displaystyle{{}##}$\hfil
      \crcr#1\crcr}}\,}
\catcode`@=12 

\begin{document}

\thesaurus{06 (02.07.1; 02.18.8; 08.02.1)}

\title{Gravitational Radiation during Thorne-\.Zytkow object formation}

\author{S.N. Nazin, K.A. Postnov}

\institute{Sternberg Astronomical Institute and  \\
Physical Department of M.V.Lomonosov State University,  Moscow, Russia\\
e-mail: nazin@sai.msu.su; pk@sai.msu.su}

\maketitle

\begin{abstract}
Calculation of gravitational radiation
during binary inspiral leading to possible formation of a Thorne-\.Zytkow (TZ)
object (a neutron star inside a supergiant core)
is performed.
The calculations were done for polytropic density distributions
with different indeces $n$, as well as for realistic
models of supergiants.
A maximum frequency
of the emitted gravitational waves during this process
was found to range from a few to 300 initial keplerian orbital
frequensies, that is
from $10^{-5}$ to $\sim 0.1$ Hz, depending on the model.
A dimensionless strain metric amplitude can reach
$h\sim 10^{-23.5}$ for a source lying 10 kpc away from the Sun.
We conclude that TZ objects forming at
a rate of 1 per 500 yrs in the Galaxy could be
potential astrophysical targets for space laser intereferometers.
Analysis of gravitational waveforms emitted during TZ-object
formation could bring a unique information about stellar
structure.


\end{abstract}

\section{Introduction}

According to the modern scenario of binary star evolution
(see van den Heuvel (1994) for a review),
massive binary systems with mass ratios far from unity
($>\sim2.5$) or in wich the mass losing star has a convective envelope
must pass through a so called common envelope (CE) stage (Paczy\'nski 1976).
Considered for the first time in detail by Taam, Bodenheimer and Ostriker
(1978), the
the double core hydrodynamical evolution in common envelope
is repeatedly calculated
(see Livio 1994 for a review). The issue of
the common envelope evolution is either ejection of part or
all of the envelope, or coalescence of the two cores with
little mass ejection leading to formation of a Thorne-\.Zytkow object
(Thorne and \.Zytkow 1977).

An important result of the extensive numerical calculations is
that the ratio of binding energy of
the stellar envelope ejected during the CE stage to change of
the system's orbital energy, $\alpha_{CE}=\Delta E_{env}/\Delta E_{orb}$
is of order unity. This implies a more than 100-fold shrinkage of
the initial orbital separation after the spiral-in for typical parameters
of high-mass X-ray binaries (HMXB)
(a supergiant with a mass $M_{sg}\sim 10-15 M_\odot$ having
a $\sim 2-3 M_\odot$
helium core in pair with a $m=1.4 M_\odot$ neutron star).
Thus, a crude lower limit for formation rate of TZ-objects can be taken
as that for HMXB, i.e.$\sim 10^{-3}$ per year per a $10^{11} M_\odot$
Milky-Way-type galaxy. In fact,
the formation rate of
TZ objects in the Galaxy typically can be nearly twice as high. This means
that $>30$ formations of TZ objects per year could be observed
from within a distance of 30 Mpc.

Let us estimate the frequency and amplitude of gravitational radiation
emitted during this process. Consider a close binary system
consisting of a neutron star orbiting a massive red (super)giant.
The dynamical friction is the main
drag  force acting the neutron star inside the supergiant's
atmosphere:
\begin{equation}
\label{force}
F_{d}(r) \sim R_G^2\rho(r) v_r(r)^2\,,
\end{equation}
where  $r$ is radial distance to the supergiant's center,
$\rho$ is its density, $v_r$ is the neutron  star velocity
relative to the envelope, and
$R_G=2Gm/(v_r^2+a_s^2)$ is gravitation capture radius for the neutron star
($G$ is the newtonian gravity constant, $a_s$ is the sound velocity).

Hydrodynamical calculations (Livio 1994) have shown that
the neutron star can reach the dense stellar core during
some $10^3$ years and the final plunge inside the dense core
takes less than one month.
A crude estimation for a dimensionless strain metric
amplitude of gravitational
radiation at the beginning of the plunge is
\begin{equation}
\label{h_est}
\eqalign{
h\sim& \frac{r_g(M)\, r_g(m)}{R_cD}\approx 5\times 10^{-26}
\left(\frac{M}{10M_\odot}\right) \times\cr
&\qquad\times\left(\frac{m}{1.4M_\odot}\right)
\left(\frac{10^{11}cm}{R_c}\right)
\left(\frac{10 Mpc}{D}\right)\,,}
\end{equation}
where $M$ and $R_c$ are mass of the star and radius of its core, respectively,
$r_g(M)=2GM/c^2$ is gravitational
radius of mass $M$ ($c$ is the speed of light)
and $D$ is a distance to the object. Obviously,
the amplitude must vanish as the neutron star approaches the
center of the star.

The maximum gravitational
radiation output during the spiral-in is
expected to occur at two-folded circular frequency of the
neutron star orbital motion. Here we will
calculate gravitational wave emission as the
neutron star plunges into a model star with
a polytropic density  distribution. We will also consider
the case of a realistic density profile in giant stars.

\section{Quadrupole gravitational radiation from\hfil\break
spiralling-in neutron star}

The transverse-traceless components of the metric perturbations
are connected with the second time derivatives of the transverse-traceless
components of the system's quadrupole moment, $\Im^{TT}_{ij}$
(see Misner, Thorne, Weeler 1973):
\begin{equation}
h^{TT}_{ij}=\frac{2G}{c^4D}\frac{d^2\Im^{TT}_{ij}}{dt^2}\,,
\end{equation}
so we need to calculate how the $\Im^{TT}_{ij}$ changes along
the neutron star trajectory.

\subsection{Quadrupole moment of the system}

We consider a point mass $m$ in a circular orbit with a
radius $r(t)<R$  inside a spherical star of mass $M$ and
radius $R$. For our model calculations we will
assume the inner structure of the giant star
not to change during the spiral-in.
Lagrangian mass comprised inside
the orbit is $M(r)=4\pi\int_0^r\rho(x)x^2dx$.

The traceless part of the second moment of the
system's mass density $\rho$ computed in a Cartesian coordinate system
centered on the binary barycenter is
\begin{equation}
\Im_{ij}(t)=\int\rho(t)(x_ix_j-\frac{1}{3}\delta_{ij}x_lx^l)d^3x\,.
\end{equation}

\begin{figure}[t]
\epsfysize=6.5cm
\vskip 6.5truecm
\caption{\label{poly_f}
Evolution of gravitational wave frequency
for different polytropic configurations. The top and right frames
are normalized for a $2 M_\odot$ core with $R=10^{10}$ cm.}
\end{figure}

We assume the binary orbit to lie in the XY coordinate plane.
We introduce a polar angle $\varphi$ between the line connecting
centres of the bodies and
OX axis. Symmetry relative to the orbital plane implies
$\Im_{xz}=\Im_{yz}=0$. Assuming the neutron star does not influences
the matter distribution inside the supergiant
(which can be the case for close enough spiralling-in
binaries, see Taam et al. (1978)) and the giant star
rotation as a whole around the barycenter,
the non-zero components of the quadrupole moment are
\begin{equation}
\Im_{xx}=(\cos^2\varphi-\frac{1}{3})\mu a^2
\end{equation}
\begin{equation}
\Im_{yy}=(\sin^2\varphi-\frac{1}{3})\mu a^2
\end{equation}
\begin{equation}
\Im_{zz}=-\frac{1}{3}\mu a^2
\end{equation}
\begin{equation}
\Im_{xy}=\frac{\sin 2\varphi}{2}\mu a^2
\end{equation}
where
\begin{equation}
\mu \equiv \frac{mM}{m+M} \hbox{ is a reduced mass}\,.
\end{equation}
This is the well-known result for two point masses
in the orbit.

\begin{figure*}[t]
\epsfxsize=\hsize
\vskip 6.5truecm
\caption{\label{TW_freq}
Evolution of the gravitational wave frequency $f_{GW}$
for Taam (1979) (a) and Weaver et al. (1979) (b)
red giant models.}
\end{figure*}

\subsection{Trajectory of the neutron star}

Equations of motion of the neutron star (treated as a point mass)
subjected to the dynamical friction force can be written as

\begin{equation}
\frac{d}{dt}\left(\frac{\partial L}{\partial \dot q^\alpha}\right) -
\frac{\partial L}{\partial q^\alpha} = -
\frac{\partial D}{\partial \dot q^\alpha}
\end{equation}

Here $L$ is Lagrange's function of the point in a gravity field
\begin{equation}
   L = \frac{m}{2}(\dot r^2 + r^2 \dot\varphi^2) + \frac{GmM(r)}{r}
\end{equation}
and $D$ is a dissipative function:
\begin{equation}
\label{dis_funct}
   D = R_G^2 \rho v^3 =
       R_G^2 \rho(r)\left[\dot r^2+r^2\dot\varphi^2\right]^{3/2}
\end{equation}
Energy losses due to dynamical friction only
are taken into account in the dissipative function.
The gravitational radiation reaction contributes unsignificantly
as it is of the order $(v/c)^5$.
We assume no change of the neutron star mass (see
Chevalier (1993) for possible issues of accretion onto spiraling-in
neutron star).

The dynamical equations were numerically solved
in order to obtain arrays of
$r(t)$, $\varphi(t)$ and their time derivatives up to the second order,
which were then used for computing the second time
derivatives of the traceless quadrupole moment $\ddot\Im_{ij}(t)$.

The density distribution inside the core was modelled by
polytripic distributions with
indeces $n$ (so that the  pressure is
$P\propto \rho^{1+1/n}$) ranging from 0 to 2.5.
We also apply our calculations
to a realistic density profile inside a 16 $M_\odot$ red giant
(Taam 1979), and to the extreme case of density distribution
inside a 15 $M_\odot$ supergiant just prior to the core collapse
(Weaver et al. 1979), which approximately corresponds to
a polytropic distribution with $n=3$.

In Fig.\ref{poly_f} we present the evolution of the two-folded orbital
frequency
(the main frequency of the gravitational wave radiation) for the
neutron star,
\begin{equation}
\eqalign{
f_{GW}&=2f_{orb}=\cr
&=6.4\times 10^{-4}(\hbox{Hz}) \left(\frac{\dot\varphi}{\pi}\right)
\sqrt{\frac{M/M_\odot}{(R/R_\odot)^3}}\,,}
\end{equation}
as a function of
time (in units of the initial keplerian time) for different
polytropes. Here $\dot\varphi$ denotes the dimensionless
time derivative.
The top and right frames are shown for $M=2 M_\odot$,
$R=10^{10}$ cm. It is clearly seen from the figure that
the larger $n$, the higher final frequency can be and
the longer time the star remains in a keplerian orbit
at the outer less denser layers. The oscillations
seen for low $n$ reflect elliptical character of the initially
circular motion suddenly subjected to a drag force. Note that for
polytropes with $n>1$ the plunge to the final frequency occurs very rapidly,
so the gravitational wave radiation will have a burst-like character for
dense configurations (see the next section).

Evolution of the gravitational wave frequency $f_{GW}$
with time for the case of density distribution inside
Taam (1979) and Weaver et al. (1979) red giants is
shown in Fig.\ref{TW_freq}.

\begin{figure}
\epsfysize=6.5cm
\vskip 6.5truecm
\caption{\label{trajec}
Trajectory of the neutron star final plunge
in a red giant with extreme density distribution
(Weaver et al. 1979).}
\end{figure}

The neutron star trajectory inside the innermost region $a<0.2 R$
for Weaver et al.'s
red giant is shown in Fig.\ref{trajec}. The duration of the final plunge is
of the order of a few 0.001 the initial keplerian time.

\subsection{Gravitational radiation}

The dimensionless waveforms $h_+$ and $h_\times$ of the metric strain
were calculated in newtonian quadrupole approximation
(see e.g. Kochanek et al 1990). As an example, we calculated
$h_+$ and $h_\times$
as seen from a distance $D$ for the face-on orientation of the binary
relative to the line of sight, so that
\begin{equation}
Dh_+=\frac{G}{c^4}(\ddot\Im_{xx}-\ddot\Im_{yy})\,;
\end{equation}
\begin{equation}
Dh_\times=2\frac{G}{c^4}\ddot\Im_{xy}\,.
\end{equation}
By measuring mass in units of the stellar mass $M$, radial distance
in units of the stellar radius $R$, time in units of the initial
free-fall time
\[
t_{ff}=\sqrt{R^3/GM}\approx 1.6\times 10^3 (\hbox{s})
\sqrt{\frac{(R/R_\odot)^3}{M/M_\odot}}\,,
\]
we can rewrite these equations in the form
\begin{equation}
\frac{D}{r_g(M)}h_+=\frac{1}{4}\frac{r_g(M)}{R}\frac{\mu}{M}
(\ddot\Psi_{xx}-\ddot\Psi_{yy})\,,
\end{equation}
\begin{equation}
\frac{D}{r_g(M)}h_\times=\frac{1}{2}\frac{r_g(M)}{R}
\frac{\mu}{M}\ddot\Psi_{xy}\,,
\end{equation}
where $\Psi_{ij}$ denotes the di\-men\-sion\-less com\-po\-nents of
the quadrupole moment $\Im_{ij}$
(cf. intuitive estimation (\ref{h_est})).

In addition, we calculated averaged over orientation of the binary
waveforms (Kochanek et al. 1990)
\begin{equation}
\eqalign{
\langle h_+^2\rangle\propto \frac{4}{15}|\ddot\Psi_{xx}&-\ddot\Psi_{zz}|^2+
\frac{4}{15}|\ddot\Psi_{yy}-\ddot\Psi_{zz}|^2+\cr
&+\frac{1}{10}|\ddot\Psi_{xx}-\ddot\Psi_{yy}|^2+
\frac{14}{15}|\ddot\Psi_{xy}|^2\,,}
\end{equation}
\begin{equation}
\langle h_\times^2\rangle\propto
\frac{1}{24}|\ddot\Psi_{xx}-\ddot\Psi_{yy}|^2+
\frac{1}{6}|\ddot\Psi_{xy}|^2\,.
\end{equation}

Temporal behaviour of these quantities for polytropes are shown
in Fig.\ref{poly_h}.
For more concentrated configurations ($n>1$),
a burst-like behaviour
corresponding to
the rapid final plunge is clearly seen.

\begin{figure}
\epsfysize=6.5cm
\vskip 6.5truecm
\caption{\label{poly_h}
Averaged over binary orientation
metric strain amplitude $\langle h^2 \rangle$
(in units of $(r_g/D)^2$) from a neutron star
spiraling-in inside a stellar core with a polytropic
density distribution, as a function of time in
units of $\protect\sqrt{R^3/GM}$.}
\end{figure}

Fig.\ref{hplusT} and \ref{hplusW} shows
the $h_+$ polarization of the gravitational wave
calculated for a face-on orientation
of the system relative to the line of sight,
and the corresponding angle averaged metric strains $\langle h^2\rangle$
for cases of Taam's and Weaver et~al's red giants.

\begin{figure*}
\epsfxsize=\hsize
\vskip 6.5truecm
\caption{\label{hplusT}
$h_+$ -- polarization of the gravitational wave (a)  and
the corresponding averaged over binary orientation
metric strain amplitude $\langle h^2 \rangle$  (b), emitted
during spiral-in of a 1.4 $M_\odot$ neutron star inside
Taam's (1979) model of red giant density distribution
($M=15 M_\odot$, $R=4.2\times 10^{12}$ cm),
as a function of time in
units of $\protect\sqrt{R^3/GM}$.}
\end{figure*}

\begin{figure*}
\epsfxsize=\hsize
\vskip 6.5truecm
\caption{\label{hplusW}
The same as in Fig.\protect\ref{hplusT} for Weaver et al.'s (1979)
density distribution inside a $16 M_\odot$ red giant with
a radius of $3.9\times 10^{13}$ cm.}
\end{figure*}

\subsection{A more realistic treatment of the
neutron star spiral-in}

We have assumed so far that the stellar structure is not
changed by the spiralling-in neutron star, which is of cause
a very crude zero approximation used in order to get simple
estimations of the effect. However, the qualitative behaviour
of the frequency and amplitude of generated gravitational
radiation is expected to not differ significantly
in more accurate calculations. The problem itself is of course
a complicated three-dimensional hydrodynamical one, but
as an illustration we performed more precise 1.5-dimensional
calculations of the neutron star spiral-in in case of
Taam's model.

We assumed after Taam (1979) that the entire common envelope
is rotating rigidly (by action of effective
convective angular momentum transfer) with
a spatially constant angular
velocity defined at each moment as $v_e=J_e r_{NS}/I_e$,
where $J_e$ is the total angular momentum transferred from the orbit
to the envelope rotation, $v_e$ the envelope circular velocity,
$r_{NS}$ the position of the neutron star and $I_e$ the moment
of inertia of the common envelope. The relative velocity of
the neutron star is $v_r=v_{NS}-v_e$. Thus, instead of
(\ref{dis_funct}) the dissipative function is
\begin{equation}
D=\frac{G^2 M_{NS}^2}{(v_r^2+a_s^2)^2} \rho(r)
v_r^3\,.
\end{equation}

The precise value of sound velocity $a_s$ is a complicated
problem and can be accurately calculated only under
fully hydrodynamical treatment. We assume after Chevalier (1993)
the motion of the neutron star in the envelope to be
mildly supersonic with a Mach number of the order
$3-1.4$ in the central regions. Therefore, during
the calculations we neglect by the sound velocity
in the outer layers of the giant, and take it to be $a_s\sim v_r$
in the inner regions after $v_r$ has been equal to the keplerian
velocity for the Lagrangian mass $M(r_{NS})$.

As expected, the calculations yielded similar dependences
of $f_{GW}$ and $h$ on time (see Fig.\ref{real_T}a, b),
with the major difference being
in time the neutron star takes to reach the final plunge.
In this case that time is much larger (as the dragging force
is much less), and is about $10^3$ years, in agreement with
the cited hydrodynamical calculations. The final plunge starts at
a distance from the center of $15\%$ the initial radius,
and lasts about several days. The GW frequency and amplitude
prove to be higher than in the simplest model case considered above
by a half an order and one order of magnitude, respectively.

\begin{figure*}
\epsfxsize=\hsize
\vskip 6.5truecm
\caption{\label{real_T}
The frequency (a) and averaged over binary orientation metric strain
amplitude $\langle h^2\rangle$ (b) of GW emitted during
spiral-in of a neutron star in the red giant model by Taam (1979)
calculated using more realistic assumptions described in the text}
\end{figure*}

\section{Discussion and conclusions}

In the present paper we calculated gravitational wave emission during
formation of a Thorne-\.Zytkow object -- a red giant star with a
neutron star core, which must be formed under certain conditions
during evolution of massive binary systems.

To estimate the gravitational wave emission, we made a number of
simplifying assumptions:
\begin{enumerate}
\item
no (or weak) influence of the spiralling-in neutron star to
the density distribution inside the giant star;
\item
the dynamical friction is the only dragging force;
\item
no accretion induced mass change of the neutron star.
\end{enumerate}

The first assumption is the most severe, as during the
spiral-in phase the whole envelope of the red giant can be lost.
However, for enough tight initial configurations (case of
Taam's model) the density profile inside the star changes
no more than an order of magnitude to the end of the spiral-in stage
(Taam 1979). This means that under more realistic conditions
the neutron star takes more time to reach the red giant's centre,
so our estimates should be considered as lower limits.
This is illustrated by calculations of the neutron star
spiral-in inside Taam's (1979) red giant model
assuming rigid rotation of the common envelope.

The second assumption seems to be justified by extensive numerical
calculations of the common envelope evolution. The third
assumption is valid while the radiation pressure prevents
matter from accreting onto the neutron star at a rate
exceeding the Eddington limit $\sim 10^{-8} M_\odot$/yr
(see discussion of another possibility in Chevalier 1993).

As is seen from Fig.\ref{poly_h}, the maximum value of the
dimensionless strain metric
amplitude $h$ during neutron star spiral-in
toward the center of a compact ($\sim R_\odot$)
polytropic configuration
can reach $\sim 10^{-24}$ at a frequency
of $\sim 10^{-2}-10^{-1}$ Hz, assuming the source
lying at 10 Mpc away from the Sun. Real density distributions
yield approximately the same amplitudes but at much
lower frequencies $\sim 5\times 10^{-5}-2\times 10^{-4}$ Hz
(Fig.\ref{TW_freq},\ref{real_T}),
which can only marginally enters into
the sensitivity range for the planned LISA detector (Laser
Interferometer Space Antenna; see Hills \& Bender (1995) for more
detail).

However, the situation is not excluded when
the neutron star enters a dilute envelope of
a 16 $M_\odot$ envelope, makes it to be dynamically
unbound but does penetrate into the inner 2-3 $M_\odot$ dense
core to form a TZ object. Then the frequency of the
GW could well fall into the range $10^{-3}-10^{-2}$,
where the LISA detector is the most sensitive.

To conclude, we stress the importance of hydrodynamical
modeling of processes considered in the present paper.
Analysis of gravitational wave signal emitted during
formation of a TZ-object can bring a unique
direct information about density distribution inside
the star, which is not available by other methods.

\section*{Acknowledgements}
The authors acknowledge Profs. Leonid Grishchuk and Kip Thorne
for stimulating dis\-cus\-sions, Prof. Vladimir Lipunov and
Dr. Mike Prokhorov for useful suggestions, and the anonymous
referee for valuable notes. The work is partially
supported by ESO CE\&E grant A-02-079, RFFI grant 94-02-04049
and INTAS grant 93-3364.

\section*{References}
\hspace\parindent

Hils, D. \& Bender, P.L., 1995, ApJ Lett. in press

Chevalier, R.A., 1993, ApJ 411, L33.

Kochanek, C.S. et al. 1990, ApJ. 358, 81.

Livio, M., 1994, in ``Interacting Binaries''
(eds. S.N.Shore, M. Livio and E.P.J. van den Heuvel), Springer: Berlin,
chapter 2.

Misner, C.W., Thorne, K.S., and Weeler, J.A., 1973, {\it Gravitation}.
W.H.Freeman \& Co.: San Francisco.

Paczy\'nski, B., 1976, in ``Structure and Evolution of Close
Binary Systems'' (eds. P.Eggeleton, S.Mitton and J.Whelan),
Reidel: Dordrecht, p.75.

Taam, R.E., 1979, ApJ Letters 20, 29.

Taam, R.E., Bodenheimer, P., and Ostriker, J.P., 1978, ApJ 222, 269.

Thorne, K.S.,  1988, in ``300 Years of Gravitaiton'', Ed. S.W.Hawking
and W.Israel, Cambridge Univ. Press.

Thorne, K.S. and \.Zytkow, A.N. 1977, ApJ 212, 832.

van den Heuvel, E.P.J., 1994, in ``Interacting Binaries''
(eds. S.N.Shore, M. Livio and E.P.J. van den Heuvel), Springer: Berlin,
chapter 3.

Weaver, T., Zimmerman, G. and Woosley, S., 1978, ApJ 225, 1021.

\end{document}